\documentstyle[preprint,eqsecnum,aps,epsf,epsfig]{revtex}
\newif\iftightenlines\tightenlinesfalse
\tightenlines\tightenlinestrue

\def\ie{{\sl i.e.}}

\def\eighth{\textstyle{1\over8}}

\iftightenlines\else\newpage\fi
\iftightenlines\global\firstfigfalse\fi
\def\dofig#1#2{\epsfxsize=#1\centerline{\epsfbox{#2}}}

\begin{document}
\draft
\preprint{\vbox{\baselineskip=14pt%
   \rightline{hep-ph/0210182}
   \rightline{UCCHEP/21-02}
}}
\title{NEUTRINO PHYSICS FROM CHARGED HIGGS AND SLEPTON 
ASSOCIATED PRODUCTION IN AMSB}
\author{Marco A. D\'\i az, Roberto A. Lineros, and Maximiliano A.
Rivera}
\author{}
\address{
Facultad de F\'\i sica, Universidad Cat\'olica de Chile,
Av. V. Mackenna 4860, Santiago, Chile
}
%
\maketitle
\begin{abstract}

In the Minimal Supersymmetric Standard Model with bilinear R-Parity 
violation, terms that violate R-Parity and lepton number are introduced 
in the superpotential, and sneutrino vacuum expectation values are 
induced. As a result, neutrino masses and mixing angles are generated 
via a low energy see-saw mechanism. We show that this model embedded 
into an anomaly mediated supersymmetry breaking scenario is testable at 
a linear collider using charged Higgs boson production in association 
with a stau. This is possible in regions of parameter space where the
charged Higgs and stau have similar mass, producing an enhancement of 
the charged scalar mixing angles. We show that the bilinear parameter
and the sneutrino vev can be determined from charged scalar observables, 
and estimate the precision of this determination.

\end{abstract}

\medskip


\newpage

%
\section{Introduction}

Over the last three decades experimental evidence has confirmed the gauge 
structure of the Standard Model (SM) with very accurate measurements. 
Nevertheless, the picture is still incomplete since the Higgs mechanism has 
not been established yet experimentally. This mechanism lies in the center 
of the mass generation problem, giving mass to the gauge bosons as well as 
the quarks and leptons. Despite this success, it is clear that the SM should 
be extended. Theoretically, the SM does not have an answer to the gauge 
hierarchy problem, nor to the stability of the Higgs mass under quantum 
corrections. Supersymmetry is one of the most popular extensions of the SM
that addresses satisfactorily these problems.

Even if supersymmetry is not a symmetry chosen by nature, there is a 
generalized feeling in the community that important discoveries are 
going to be available soon after the completion of the new generation of 
colliders, starting with the Large Hadron Collider (LHC) at CERN, or maybe 
even at the already running Tevatron at Fermilab. A Linear 
Collider (LC) is crucial in order to study the new phenomena and its 
relation to physics at even higher energy scales. This relation is more 
than a simple complementarity with the LHC, since the ideal is for them to
run simultaneously in such a way that the discoveries in one machine may
influence the running parameters of the other\cite{LC}.

In parallel to the physics we can learn form colliders goes neutrino
physics. Today, neutrino physics has become one of the most exiting areas
in particle physics, with experimental results that indicate neutrinos
have a mass and oscillate. Atmospheric neutrino data indicates a 
$\nu_{\mu}-\nu_{\tau}$ mixing with an angle $0.3<\sin^2\theta_{atm}<0.7$
and a mass splitting of $0.03<\sqrt{\Delta m^2_{atm}}<0.07$ eV. Solar 
neutrino data favors a LMA solution with the best fit given by 
$\tan^2\theta_{sol}=0.44$ and $\sqrt{\Delta m^2_{sol}}=0.008$ 
eV\cite{valleetal}. There is little doubt now that neutrinos have mass
and this is the first experimental evidence that the SM must be modified.

Supersymmetry with Bilinear R-Parity Violation 
(BRpV)\cite{BRpVrecent,mnu1lshort,numassBRpV,BRpVothers,BRpV_tau,BRpVmore} 
is a model in which 
neutrino masses and mixing angles are generated by the presence of 
bilinear terms in the superpotential which violate R-Parity and lepton 
number. At the same time, sneutrino vacuum expectation values are induced.
In this model, neutralinos mix with neutrinos, generating at tree 
level a mass for one of the neutrinos, while the other two acquire a 
mass at one loop. 

One of the nicest features of supersymmetric models which violates
R-Parity and lepton numbers through BRpV is that neutrino physics is 
closely related to high energy
physics in such a way that neutrino properties can be tested at future
colliders, in particular at a LC where it is possible to make precision 
measurements on different observables. It is already understood that 
neutrino mixing angles are related to ratios of branching ratios of 
the neutralino, which being the Lightest Supersymmetric Particle (LSP), 
will have non-suppressed R-Parity violating branching 
ratios\cite{mnu1lshort,LSPdecay}.

In BRpV there is also mixing in the scalar sector, in particular, charged 
Higgs fields mix with charged sleptons. As a result, in electron positron 
collisions it is possible to produce a charged Higgs boson in association 
with a stau, for example. This kind of R-Parity violating processes
are not present in trilinear R-Parity violating models and, therefore, 
would be a signature of BRpV. In this paper we study charged Higgs 
production in association with a stau and its relation to neutrino physics
in a model with Anomaly Mediated Supersymmetry Breaking 
(AMSB)\cite{AnomMed,others,phenom},
where an enhancement of charged scalar mixing angles can occur due to
mass degeneracy between the scalars\cite{brazil}.
We study how parameters relevant to solar neutrino physics, like the
$\epsilon_i$ parameters in the superpotential, can be extracted from collider 
observables in the charged scalar sector.

\section{Bilinear R-Parity Violation and Neutrino Physics}

In the BRpV--MSSM model, explicit lepton number and R--Parity violating 
terms are added to the MSSM superpotential
\begin{equation}
W=W_{MSSM}+\epsilon_i\widehat L_i\widehat H_u
\end{equation}
where the three $\epsilon_i$ parameters have unit of mass. In addition, 
corresponding soft terms add included in the lagrangian
\begin{equation}
{\cal L}_{soft}={\cal L}_{soft}^{MSSM}+B_i\epsilon_i\widetilde L_i H_u
\end{equation}
with $B_i$ being the bilinear soft terms associated to $\epsilon_i$.
These terms induce sneutrino vacuum expectation values $v_i$ which 
contribute to the gauge boson masses. Our notation for the fields that 
acquire a vev is
\begin{equation} 
H_d={{H^0_d}\choose{H^-_d}}\,,\qquad 
H_u={{H^+_u}\choose{H^0_u}}\,,\qquad
\widetilde L_i={{\tilde L^0_i}\choose{\tilde\ell^-_i}}\,,
\label{eq:shiftdoub} 
\end{equation} 
where
\begin{equation}
H_d^0\equiv{1\over{\sqrt{2}}}[\sigma^0_d+v_d+i\varphi^0_d]\,,\quad
H_u^0\equiv{1\over{\sqrt{2}}}[\sigma^0_u+v_u+i\varphi^0_u]\,,\quad
\tilde{L}_i^0\equiv{1\over{\sqrt{2}}}[\tilde\nu^R_i+v_i+i\tilde\nu^I_i]\,.
\label{shiftedfields}
\end{equation}
The tree level scalar potential contains the following linear terms 
\begin{equation}
V_{linear}^0=t_d^0\sigma^0_d+t_u^0\sigma^0_u+t_1^0\tilde\nu^R_1
+t_2^0\tilde\nu^R_2+t_3^0\tilde\nu^R_3\,,
\label{eq:Vlinear}
\end{equation}
where the different $t^0$ are the tadpoles at tree level. They are given by
\begin{eqnarray}
t_d^0&=&\Big(m_{H_d}^2+\mu^2\Big)v_d+v_dD-\mu\Big(Bv_u+v_j\epsilon_j\Big)
\nonumber\\
t_u^0&=&-B\mu v_d+\Big(m_{H_u}^2+\mu^2\Big)v_u-v_uD+v_jB_j\epsilon_j
+v_u\epsilon^2
\label{eq:tadpoles}\\
t_i^0&=&v_iD+\epsilon_i\Big(-\mu v_d+v_uB_i+v_j\epsilon_j\Big)+
v_iM^2_{Li}
\nonumber
\end{eqnarray}
where we have defined $D=\eighth(g^2+g'^2)(v_1^2+v_2^2+v_3^2+v_d^2-v_u^2)$, 
$\epsilon^2=\epsilon_1^2+\epsilon_2^2+\epsilon_3^2$, and there is a sum 
over $j$ but not over $i$. The five tree level tadpoles $t_{\alpha}^0$ are 
equal to zero at the minimum of the tree level potential, and from there 
one can determine the five tree level vacuum expectation values.

As a consequence of the $\epsilon$ terms and the sneutrino vevs R-Parity
and lepton number are violated, and R-Parity even fields mix with R-Parity 
odd fields. The most conspicuous of these is the mixing between neutralinos 
and neutrinos. This is important because due to a low energy see-saw 
mechanism, the neutrinos acquire mass.
\begin{equation}
m_{eff}=
\frac{M_1 g^2 \!+\! M_2 g'^2}{4\, det({\cal M}_{\chi^0})} 
\left[
\matrix{
\Lambda_e^2 & \Lambda_e \Lambda_\mu & \Lambda_e \Lambda_\tau \cr
\Lambda_e \Lambda_\mu & \Lambda_\mu^2 & \Lambda_\mu \Lambda_\tau \cr
\Lambda_e \Lambda_\tau & \Lambda_\mu \Lambda_\tau & \Lambda_\tau^2
}\right]
\label{meff}
\end{equation}
where ${\cal M}_{\chi^0}$ is the $4\times4$ neutralino mass matrix, and 
we define 
\begin{equation}
\Lambda_i=\mu v_i+\epsilon_i v_d\,.
\end{equation}
These important parameters define the neutrino physics at tree level. The
effective neutrino mass matrix in eq.~(\ref{meff}) has only one eigenvalue 
different from zero, and it is given by
\begin{equation}
\label{mnutree}
\sqrt{\Delta m_{atm}^2}\approx m_{\nu_3}= 
\frac{M_1 g^2 + M_2 g'^2}{4\, det({\cal M}_{\chi^0})} 
|{\vec \Lambda}|^2.
\end{equation}
Since the lowest two eigenvalues are massless and degenerate, there is no 
meaningful solar angle at tree level. On the other hand, the near maximal 
atmospheric angle is 
\begin{equation}
\tan\theta_{23}=\Lambda_{\mu}/\Lambda_{\tau}\approx 1
\end{equation}
and the Chooz constraint $\sin^2\theta_{13}<0.045$\cite{Chooz} can be 
satisfied taking
\begin{equation}
|\tan\theta_{13}|=|\Lambda_e|/\sqrt{\Lambda_{\tau}^2+\Lambda_{\mu}^2}
\ll 1\,.
\end{equation}
Information on the solar neutrino mass and angle can be obtained only when
one-loop corrections are added to the neutrino mass matrix. As an example, 
the bottom/sbottom loops give a simple contribution which is approximately
given by
\begin{equation}
\sqrt{\Delta m_{sol}^2}\approx m_{\nu_2}\sim{{3h_b^2|\vec\epsilon|^2}
\over{16\pi^2}}{{m_b}\over{M_{SUSY}^2}}\ln{{M_{\tilde b_2}^2}\over
{M_{\tilde b_1}^2}}\,.
\end{equation}
In supergravity models, this is the most important loop, followed by 
charged Higgs loops. It is clear from this formula that the $\epsilon$
parameters have an important effect in solar neutrino physics, as opposed 
to the $\Lambda$ parameters which are important for atmospheric neutrino 
physics. 

When details of the neutrino physics are not the main issue, it has been 
proven very useful to work in the approximation where BRpV is introduced 
only in one generation, say the stau. In this case, the atmospheric scale 
still is given by eq.~(\ref{mnutree}). In the rest of this article we will 
follow this approach.

\section{Benchmark for BRpV-AMSB}

In order to study the effects of BRpV in an AMSB scenario, we have
chosen a case study, or benchmark, in which we find non negligible
charged Higgs production in association with a stau. The parameters
which define our model are
\begin{eqnarray}
M_{3/2}&=&30 \,\,{\mathrm TeV} 
\nonumber\\
m_0&=&200 \,\,{\mathrm GeV}
\nonumber\\
\tan\beta&=&15
\nonumber\\
\mu&<&0
\label{benchmark}\\
\epsilon_3&=&1 \,\,{\mathrm GeV}
\nonumber\\
m_{\nu}&=&0.1 \,\,{\mathrm eV}
\nonumber
\end{eqnarray}
The value of $\mu$ is fixed by imposing the correct electroweak
symmetry breaking to be $\mu=-466$ GeV. The neutrino mass fix the
sneutrino vacuum expectation value to $v_3=0.035$ GeV. The values of
$B$ and $B_3$ are also determined by the tadpole equations.

Regarding the spectrum of this model, the LSP is the lightest
neutralino, with a mass $m_{\tilde\chi^0_1}=94.39$ GeV, followed closely
by the lightest chargino with a mass $m_{\tilde\chi^+_1}=94.43$ GeV.
In the neutral scalar sector we have a CP-odd Higgs with $m_A=171$
GeV, and a tau-sneutrino with $m_{\tilde\nu_{\tau}}=142$ GeV. It
should be stressed that in BRpV models the CP-odd Higgs boson mixes
with the sneutrinos. We here call the CP-odd Higgs to the eigenstate
with largest component to the original Higgs fields.

In this article we are interested in the charged scalar sector. The
charged Higgs fields and the left and right staus mix to form a set of 
four charged scalars, one of them being the charged Goldstone boson. 
Among the other three eigenstates we call charged Higgs to the one with
largest component to the original charged Higgs fields. The charged
scalar spectrum is
\begin{eqnarray}
m_{H^+}&=&188 \,\,{\mathrm GeV}
\nonumber\\
m_{\tilde\tau^+_1}&=&115 \,\,{\mathrm GeV}
\label{chscalarmasses}\\
m_{\tilde\tau^+_2}&=&190 \,\,{\mathrm GeV}
\nonumber
\end{eqnarray}
The $4\times4$ charged scalar mass matrix is diagonalized by a
$4\times4$ rotation matrix, which for our benchmark is given by
\begin{equation}
{\bf R}=\left[\matrix{
-0.067 &  0.998 & -0.0001 &  0    \cr
-0.80  & -0.05  &  0.45   &  0.40 \cr
-0.014 & -0.001 &  0.65   & -0.76 \cr
 0.60  &  0.04  &  0.62   &  0.51}\right]
\label{Rmatrix}
\end{equation}
where the columns correspond to the fields 
$(H_u^+,H_d^+,\tilde\tau_L^+,\tilde\tau_R^+)$ and the rows to 
$(G^+,H^+,\tilde\tau_1^+,\tilde\tau_2^+)$. From this rotation matrix
we learn that the Goldstone boson has no right handed stau component and
very little left handed stau component, as it should be. We also see
that the light stau has almost no component to the Higgs fields, \ie,
it is almost pure stau, and that the charged Higgs has an important 
component of stau.

\section{Charged Higgs/Slepton Sector}

As we mentioned, in BRpV charged Higgs bosons and charged sleptons
mix forming, in the general three generations case, a $8\times 8$ mass 
matrix, and in the simplified case of BRpV only in one generation, a
$4\times 4$ mass matrix. The relevant mass terms in the scalar
potential are
\begin{equation}
V_{quadratic}=\left[H_u^-,H_d^-,\tilde\tau_L^-,\tilde\tau_R^-\right]
{\bf M}_{S^{\pm}}^2
\left[\matrix{H_u^+ \cr H_d^+ \cr \tilde\tau_L^+ \cr \tilde\tau_R^+}
\right]
\end{equation}
In the R-Parity conserving limit (MSSM), the mass matrix is diagonal in
$2\times 2$ blocks, and the charged Higgs sector is decoupled from the
charged slepton sector. We write the mass matrix in the following form
\begin{equation}
{\bf M}_{S^{\pm}}^2={\bf M}_{S^{\pm}}^{2(0)}+{\bf M}_{S^{\pm}}^{2(1)}
\label{completeM}
\end{equation}
motivated by the fact that BRpV terms are small. The MSSM part is
\begin{equation}
{\bf M}_{S^{\pm}}^{2(0)}=\left[\matrix{
m_{H^{\pm}}^{2(0)}s_{\beta}^2 & 
m_{H^{\pm}}^{2(0)}s_{\beta}c_{\beta} & 0 & 0 \cr
m_{H^{\pm}}^{2(0)}s_{\beta}c_{\beta} & 
m_{H^{\pm}}^{2(0)}c_{\beta}^2 & 0 & 0 \cr
0 & 0 & \widehat M_{L_3}^2 & \widehat M_{LR}^2 \cr
0 & 0 & \widehat M_{LR}^2 & \widehat M_{R_3}^2
}\right]
\label{mat8x8}
\end{equation}
where the diagonal slepton mass entries are given by
\begin{eqnarray}
\widehat M_{L_3}^2&=&M_{L_3}^2-\textstyle{1\over8}(g^2-g'^2)
(v_d^2-v_u^2)+m_{\tau}^{2(0)}\,,
\nonumber\\
\widehat M_{R_3}^2&=&M_{R_3}^2-\textstyle{1\over4}g'^2(v_d^2-v_u^2)+
m_{\tau}^{2(0)}\,,
\\
\widehat M_{LR}^2&=&m_{\tau}^{(0)}(A_{\tau}-\mu\tan\beta)
\nonumber
\end{eqnarray}
and $m_{\tau}^{(0)}$ is the tau lepton mass calculated in the MSSM
approximation. 
Note that since in our model the tau mixes with the
charginos, the physical tau mass has a more complicated dependence on
the parameters of the model\cite{Akeroydetal}. 
We write the BRpV contributions to the charged scalar mass matrix as
\begin{equation}
{\bf M}_{S^{\pm}}^{2(1)}=\left[\matrix{
\Delta m_{H_u}^2 & 0 & X_{uL} & X_{uR} \cr
0 & \Delta m_{H_d}^2 & X_{dL} & X_{dR} \cr
X_{uL} & X_{dL} & \Delta m_{L_3}^2 & 0 \cr
X_{uR} & X_{dR} & 0 & \Delta m_{R_3}^2
}\right]
\label{M1}
\end{equation}
with the diagonal contributions given by:
\begin{eqnarray}
\Delta m_{H_u}^2&=&\mu\epsilon_3\textstyle{{v_3}\over{v_d}}
-\textstyle{1\over4}g^2v_3^2+\textstyle{1\over2}h_{\tau}^2v_3^2\,,
\nonumber\\
\Delta m_{H_d}^2&=&\textstyle{{v_3^2}\over{v_d^2}}
\textstyle{{c_{\beta}^2}\over{s_{\beta}^2}}\overline m_{\tilde\nu}^2
-\mu\epsilon_3\textstyle{{v_3}\over{v_d}}
\textstyle{{c_{\beta}^2}\over{s_{\beta}^2}}
+\textstyle{1\over4}g^2v_3^2\,,
\\
\Delta m_{L_3}^2&=&\epsilon_3^2+\textstyle{1\over8}g_Z^2v_3^2\,,
\nonumber\\
\Delta m_{R_3}^2&=&\textstyle{1\over2}h_{\tau}^2v_3^2+
\textstyle{1\over4}g'^2v_3^2\,,
\nonumber
\end{eqnarray}
where $\overline m_{\tilde\nu}^2=m_{\tilde\nu}^{(0)2}+\epsilon_3^2
+{1\over8}g_Z^2v_3^2$ and $g_Z^2\equiv g^2+g'^2$. The quantity
$m_{\tilde\nu}^{(0)}$ is the sneutrino mass in the 
$\epsilon_3\rightarrow 0,v_3\rightarrow 0$ limit\cite{brazil}.
The off-diagonal BRpV terms of in the charged scalar matrix in 
eq.~(\ref{M1}) are
\begin{eqnarray}
X_{uL}&=&\textstyle{1\over4}g^2v_dv_3-\mu\epsilon_3-\textstyle{1\over2}
h_{\tau}^2v_dv_3\,,
\nonumber\\
X_{uR}&=&-\textstyle{1\over{\sqrt{2}}}h_{\tau}(A_{\tau}v_3+\epsilon_3v_u)\,,
\nonumber\\
X_{dL}&=&{{v_3}\over{v_d}}{{c_{\beta}}\over{s_{\beta}}}
\overline m_{\tilde\nu}^2-\mu\epsilon_3{{c_{\beta}}\over{s_{\beta}}}
+\textstyle{1\over4}g^2v_uv_3\,,
\\
X_{dR}&=&-\textstyle{1\over{\sqrt{2}}}h_{\tau}(\mu v_3+\epsilon_3v_d)\,.
\nonumber
\end{eqnarray}
The complete charged scalar mass matrix in eq.~(\ref{completeM}) has a
zero eigenvalue corresponding to the charged Goldstone boson
$G^{\pm}$. This eigenvalue can be isolated with the rotation
\begin{equation}
{\bf R}_G=\left[\matrix{
c_{\beta}r & -s_{\beta}r & -\textstyle{{v_3}\over{v_d}}c_{\beta}r & 0 \cr
s_{\beta} & c_{\beta} & 0 & 0 \cr
-\textstyle{{v_3}\over{v_d}}c_{\beta}^2r & 
\textstyle{{v_3}\over{v_d}}s_{\beta}c_{\beta}r & r & 0 \cr
0 & 0 & 0 & 1
}\right]
\label{rotGolds}
\end{equation}
where we have defined the factor
\begin{equation}
r={1\over{\sqrt{1+{{v_3^2}\over{v_d^2}}c_{\beta}^2}}}\,.
\end{equation}

\begin{figure}
\dofig{5in}{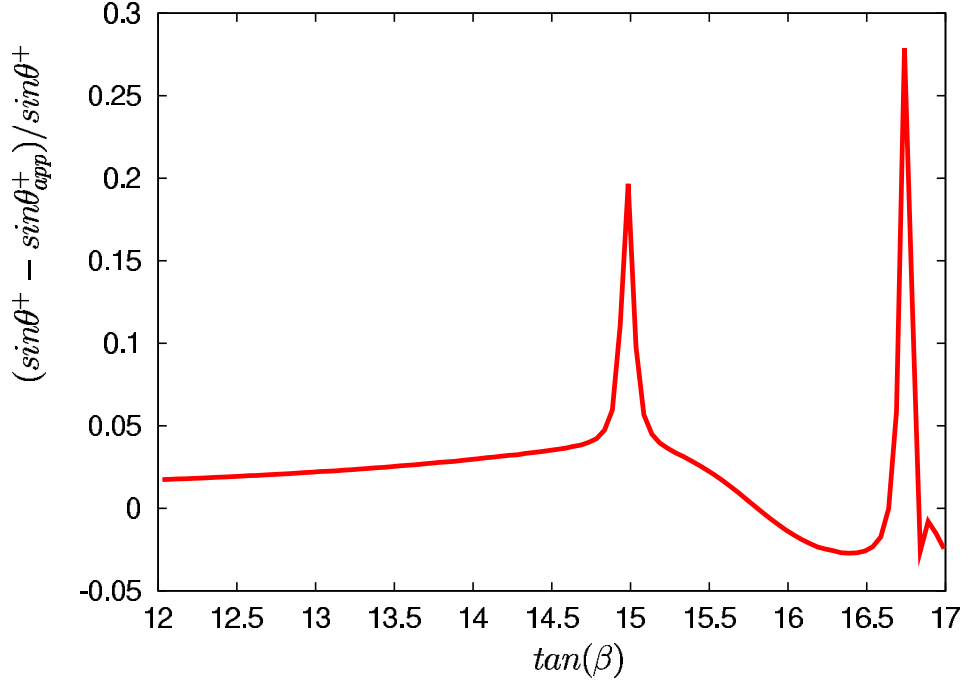}
\vskip .5cm
\caption[]{
The relative error of the approximated formula for the angle 
$\sin\theta^+$, which indicates how much component of stau fields has the 
charged Higgs, is plotted against $\tan\beta$.
}
\label{theta+app}
\end{figure}

In what follows we present some approximated formulas for masses and 
mixing angles between charged Higgs bosons and charged scalars. Our aim
is to find second order corrections to the masses and first order mixing 
angles. In order to do that, it is important to perform first the rotation 
defined in eq.~(\ref{rotGolds}), otherwise, approximating over the 
unrotated $4\times4$ matrix in eq.~(\ref{mat8x8}) would introduce 
fictitious corrections to the zero Goldstone boson mass. In addition we 
rotate the stau sector by an angle $\theta_{\tilde\tau}$:
\begin{equation}
{\bf R}_{\tilde\tau}=\left[\matrix{
1 & 0 & 0 & 0 \cr
0 & 1 & 0 & 0 \cr
0 & 0 & c_{\tilde\tau} & s_{\tilde\tau} \cr
0 & 0 & -s_{\tilde\tau} & c_{\tilde\tau}
}\right]
\end{equation}
after which the zeroth order mass matrix becomes diagonal:
\begin{equation}
{\bf R}_{\tilde\tau}{\bf R}_G{\bf M}_{S^{\pm}}^2
{\bf R}_G^T{\bf R}_{\tilde\tau}^T=
\left[\matrix{0 & 0 & 0 & 0 \cr
0 & m_{H^{\pm}}^{2(0)}+\Delta\hat m_{H^{\pm}}^2 & X_{H\tilde\tau_1} & 
X_{H\tilde\tau_2} \cr
0 & X_{H\tilde\tau_1} & m_{\tilde\tau_1}^{2(0)} + \Delta\hat 
m_{\tilde\tau_1}^2 & 0 \cr
0 & X_{H\tilde\tau_2} & 0 & m_{\tilde\tau_2}^{2(0)} + \Delta\hat 
m_{\tilde\tau_2}^2
}\right]
\end{equation}
The corrections to the diagonal elements up to second order are
\begin{eqnarray}
\Delta\hat m_{H^{\pm}}^2&=&s_{\beta}^2\Delta m_{H_u}^2+
c_{\beta}^2\Delta m_{H_d}^2\,,
\nonumber\\
\Delta\hat m_{\tilde\tau_1}^2&=&c_{\tilde\tau}^2\Delta m_{L_3}^2+
s_{\tilde\tau}^2\Delta m_{R_3}^2
+\textstyle{{v_3^2}\over{v_d^2}}c_{\beta}^2c_{\tilde\tau}\left(
c_{\tilde\tau}\widehat M_{L_3}^2+s_{\tilde\tau}\widehat M_{LR}^2\right)\,,
\\
\Delta\hat m_{\tilde\tau_2}^2&=&s_{\tilde\tau}^2\Delta m_{L_3}^2+
c_{\tilde\tau}^2\Delta m_{R_3}^2+
\textstyle{{v_3^2}\over{v_d^2}}c_{\beta}^2s_{\tilde\tau}\left(
s_{\tilde\tau}\widehat M_{L_3}^2-c_{\tilde\tau}\widehat M_{LR}^2\right)
\,.
\nonumber
\end{eqnarray}
and the off diagonal corrections up to first order are
\begin{eqnarray}
X_{H\tilde\tau_1}&=&c_{\tilde\tau}\left(s_{\beta}X_{uL}+
c_{\beta}X_{dL}\right)+s_{\tilde\tau}\left(s_{\beta}X_{uR}+
c_{\beta}X_{dR}\right)\,,
\nonumber\\
X_{H\tilde\tau_2}&=&-s_{\tilde\tau}\left(s_{\beta}X_{uL}+
c_{\beta}X_{dL}\right)+c_{\tilde\tau}\left(s_{\beta}X_{uR}+
c_{\beta}X_{dR}\right)\,.
\end{eqnarray}
The last mixings $X_{H\tilde\tau_1}$ and $X_{H\tilde\tau_2}$ are rotated 
away with the aid of two small rotations
\begin{equation}
{\bf R}_1=\left[\matrix{
1 &  0   &  0  & 0 \cr
0 &  c_1 & s_1 & 0 \cr
0 & -s_1 & c_1 & 0 \cr
0 &  0   &  0  & 1 }\right]\,\,,\qquad
{\bf R}_2=\left[\matrix{
1 &   0  & 0 &  0  \cr
0 &  c_2 & 0 & s_2 \cr
0 &   0  & 1 &  0  \cr
0 & -s_2 & 0 & c_2 }\right]\,\,.
\end{equation}
with rotation angles given by
\begin{equation}
s_1={{X_{H\tilde\tau_1}}\over{m_{H^{\pm}}^{2(0)}-m_{\tilde\tau_1}^{2(0)}}}
\,,\qquad
s_2={{X_{H\tilde\tau_2}}\over{m_{H^{\pm}}^{2(0)}-m_{\tilde\tau_2}^{2(0)}}}
\,.
\end{equation}
These are the mixing angles that are potentially large when the Higgs 
and stau masses are nearly degenerate. The final rotation that
diagonalizes the charged Higgs/stau mass matrix is given by
\begin{equation}
{\bf R}={\bf R}_1{\bf R}_2{\bf R}_{\tilde\tau}{\bf R}_G
\end{equation}
It is clear that these approximations will fail in the case that two 
eigenvalues are nearly degenerate. We call
\begin{equation}
\sin^2\theta^+={\bf R}_{23}^2+{\bf R}_{24}^2
\end{equation}
which indicates how much of stau fields has the charged Higgs. In 
Fig.~\ref{theta+app} we compare the approximated formula with the exact 
numerical value of $\sin\theta^+$. The relative error is plotted against
$\tan\beta$ for fixed values of $M_{3/2}=30$ TeV, $m_0=200$ GeV, 
$\epsilon_3=1$ GeV, $m_{\nu}=0.1$ eV, and $\mu<0$. The error stays within
$5\%$ except for the points of near degeneracy. Close to $\tan\beta=15$
the charged Higgs and heavy stau masses are similar, and for 
$\tan\beta\approx 17$ the charged Higgs and light stau masses are similar.

The second order corrections to the Higgs and stau masses are
\begin{eqnarray}
\Delta m_{H^{\pm}}^2&=&\Delta\hat m_{H^{\pm}}^2+s_1X_{H\tilde\tau_1}+
s_2X_{H\tilde\tau_2}\,,
\nonumber\\
\Delta m_{\tilde\tau_1}^2&=&\Delta\hat m_{\tilde\tau_1}^2-
s_1X_{H\tilde\tau_1}\,,
\\
\Delta m_{\tilde\tau_2}^2&=&\Delta\hat m_{\tilde\tau_2}^2-
s_2X_{H\tilde\tau_2}\,.
\nonumber
\end{eqnarray}
and since they are of second order in R-Parity violating parameters, they 
are small, although they may be enhanced if there is near degeneracy.

\section{Charged Higgs and Stau Associated Production}

A couple of charged scalars are produced in electron--positron 
annihilation via the interchange in the s--channel of a photon and a 
$Z$ boson. The total production cross section is
\begin{equation}
\sigma(e^+e^-\rightarrow S_i^+S_j^-)={{\lambda^{3/2}}\over{24\pi s}}
\Bigg[
{{e^4}\over 2}\delta_{ij}-{{ge^2g_V^e}\over{2c_W}}P_Z\delta_{ij}
\lambda_{ZSS}^{ij}+{{g^2(g_V^{e2}+g_A^{e2})}\over{8c_W^2}}P_Z^2
\lambda_{ZSS}^{ij2}\Bigg]
\end{equation}
with $g_V^e={1\over2}-2s_W^2$, $g_A^e={1\over2}$, and $P_Z=s/(s-m_Z^2)$.
In the case of mixed production, only the $Z$ boson contributes, with a 
strength determined by the $ZS^+S^-$ coupling, $\lambda_{ZSS}^{ij}$. In
the unrotated basis $(H_u,H_d,\tilde\tau_L,\tilde\tau_R)$ these couplings
are
\begin{equation}
{\bf \lambda}'_{ZSS}={g\over{2c_W}}\left[\matrix{
-c_{2W} & 0 & 0 & 0 \cr
0 & -c_{2W} & 0 & 0 \cr
0 & 0 & -c_{2W} & 0 \cr
0 & 0 & 0 & 2s_W^2    }\right]
\end{equation}
Since $s_W^2\approx1/4$ the couplings are approximately proportional
to ${\bf \lambda}'_{ZSS}\sim\rm diag(1,1,1-1)$. The rotated couplings
are of course ${\bf \lambda}_{ZSS}={\bf R}{\bf \lambda}'_{ZSS}{\bf R}^T$.

For the chosen benchmark, the non negligible production cross sections
are given by,
\begin{eqnarray}
\sigma(e^+e^-\rightarrow \tilde\tau_1^+\tilde\tau_1^-)&=22.9
\,\,{\mathrm fb}
& \nonumber\\
\sigma(e^+e^-\rightarrow H^+H^-)&=22.2 \,\,{\mathrm fb}
& \nonumber\\
\sigma(e^+e^-\rightarrow \tilde\tau_2^+\tilde\tau_2^-)&=21.0 
\,\,{\mathrm fb}
& \nonumber\\
\sigma(e^+e^-\rightarrow \tilde\tau_1^{\pm}\tilde\tau_2^{\mp})&=3.44 
\,\,{\mathrm fb}
& \label{CrossSections}\\
\sigma(e^+e^-\rightarrow H^{\pm}\tilde\tau_1^{\mp})&=1.81 
\,\,{\mathrm fb}
& \nonumber\\
\sigma(e^+e^-\rightarrow H^{\pm}\tilde\tau_2^{\mp})&=0.79 
\,\,{\mathrm fb}
& \nonumber
\end{eqnarray}
where the last two violate R-Parity. With a projected integrated luminosity 
of ${\cal L}=500 \,\,{\mathrm fb}^{-1}$, we expect plenty of events where
charged Higgs are produced in association with staus. These cross sections 
are governed by the couplings ${\bf \lambda}_{ZSS}^{ij}$, which for our 
benchmark are approximately
\begin{equation}
{\bf \lambda}_{ZSS}\approx{g\over{2c_W}}\left[\matrix{
-0.5 &  0    &  0    &  0    \cr
 0   & -0.24 & -0.39 &  0.2  &  \cr
 0   & -0.39 &  0.08 & -0.3  &  \cr
 0   &  0.2  & -0.3  & -0.34 &  }\right]
\end{equation}
written in the base $(G^+,H^+,\tilde\tau_1^+,\tilde\tau_2^+)$.
The first thing to notice is that the $Z$ couples to a pair of light staus 
with a strength 3 to 4 times smaller than to the other two scalars. The 
reason is that the light stau is almost pure stau and has nearly equal 
left and right component, and the minus sign in ${\bf \lambda}'_{ZSS}$ 
induces a cancellation. For this reason its cross section is comparable 
to the other two pair productions even though there is more phase space 
for light staus. The same type of cancellation does not happen to the heavy 
stau because this eigenstate has also a large component of Higgs fields.

The second thing worth noticing is that despite the fact the light stau is 
almost pure stau, as indicated by eq.~(\ref{Rmatrix}), its mixed production 
together with a charged Higgs is larger than the mixed production of a 
heavy stau together with a charged Higgs, as shown in 
eq.~(\ref{CrossSections}). The explanation is simple: the charged Higgs 
has a large component of stau. 

\begin{figure}
\dofig{5in}{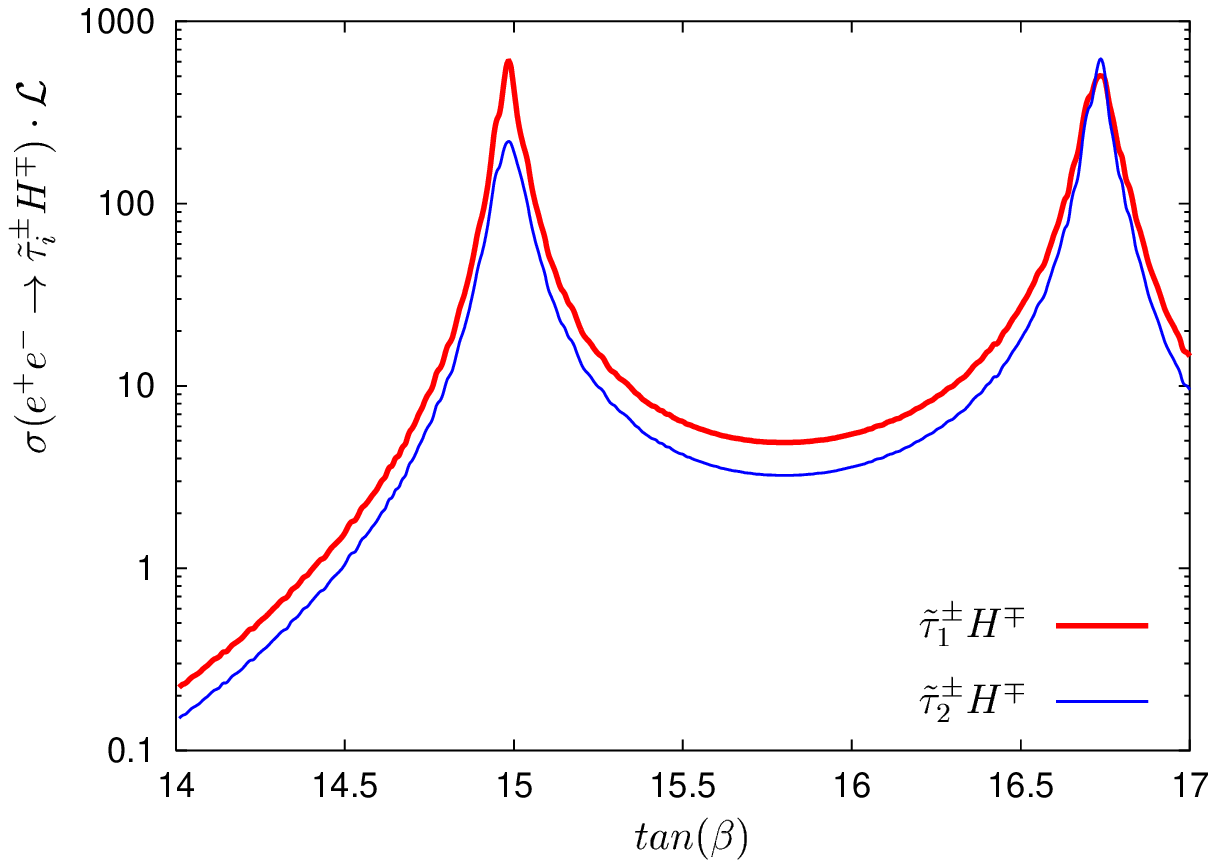}
\vskip .5cm
\caption[]{
Charged Higgs production in association with a stau multiplied by an 
integrated luminosity of $500 \,\,{\mathrm fb}^{-1}$, plotted as a function 
of $\tan\beta$, for $M_{3/2}=30$ TeV, $m_0=200$ GeV, $\epsilon_3=1$ GeV,
$m_{\nu}=0.1$ eV, and $\mu<0$.
}
\label{Stanb}
\end{figure}

In Fig.~\ref{Stanb} we plot as a function of $\tan\beta$ the expected number 
of events in which a charged Higgs is produced in association with a stau, 
where we have summed over both possible signs of the scalar electric charge.
The cross section has a maximum near $\tan\beta=15$ where the charged Higgs 
mass is very similar to the mass of the heavy stau. The charged Higgs
associated production with a light stau also has a maximum at
$\tan\beta\approx15$ because at this point the charged Higgs has a large 
stau component, as explained before. As we increase $\tan\beta$ the
charged Higgs mass decreases and eventually, near $\tan\beta\approx16.7$,
becomes similar to the light stau mass, increasing again the cross section.

\begin{figure}
\dofig{5in}{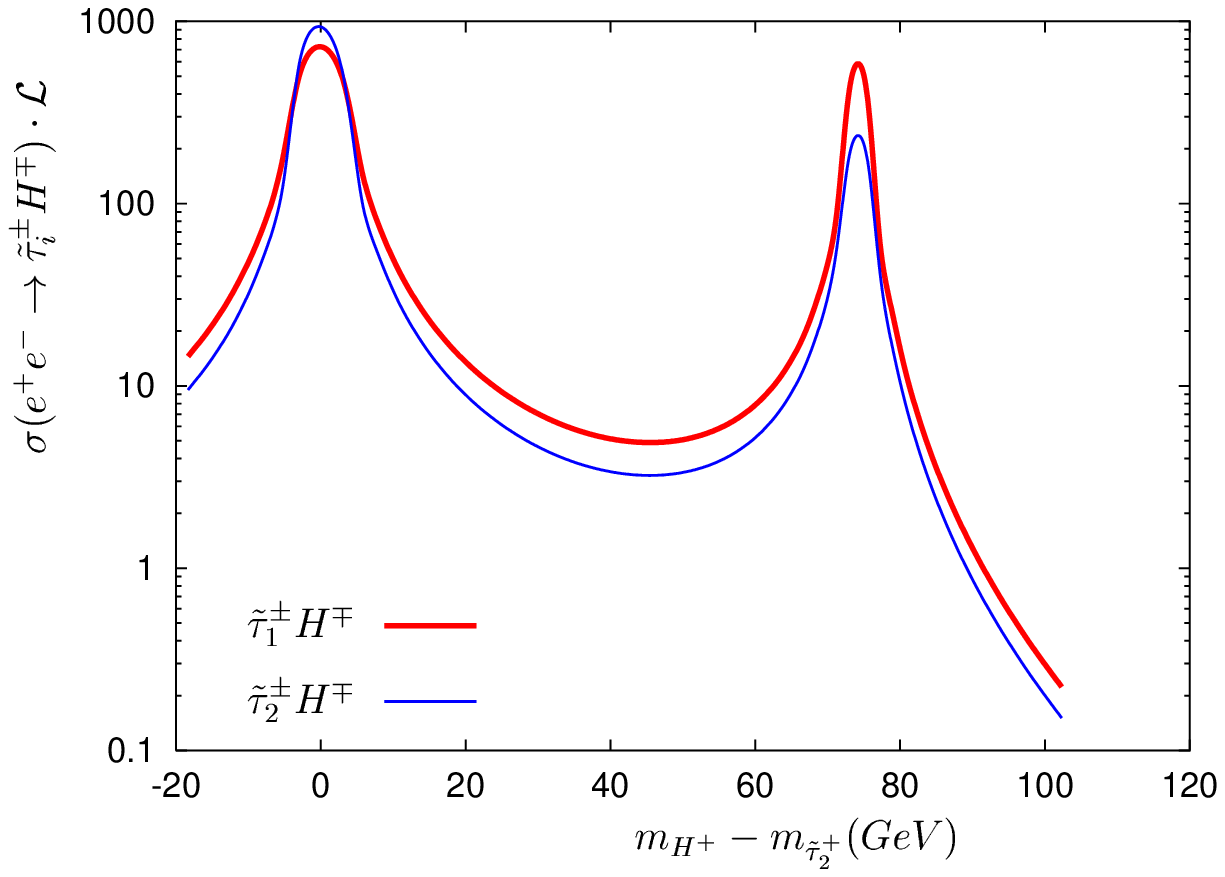}
\vskip .5cm
\caption[]{
Charged Higgs production in association with a stau multiplied by an 
integrated luminosity of $500 \,\,{\mathrm fb}^{-1}$, plotted as a function 
of the mass splitting $m_{H^+}-m_{\tilde\tau_2}$, 
for $M_{3/2}=30$ TeV, $m_0=200$ GeV, $\epsilon_3=1$ GeV,
$m_{\nu}=0.1$ eV, and $\mu<0$.
}
\label{SdeltaM}
\end{figure}

In order to have an idea of the degree of degeneracy needed for large 
cross sections, we plot in Fig.~\ref{SdeltaM} a similar graph but this time 
as a function of the mass splitting $m_{H^+}-m_{\tilde\tau_2}$. A mass
difference of 10 GeV gives of the order of 50 events, and increasing 
exponentially with decreasing mass difference.

\begin{figure}
\dofig{5in}{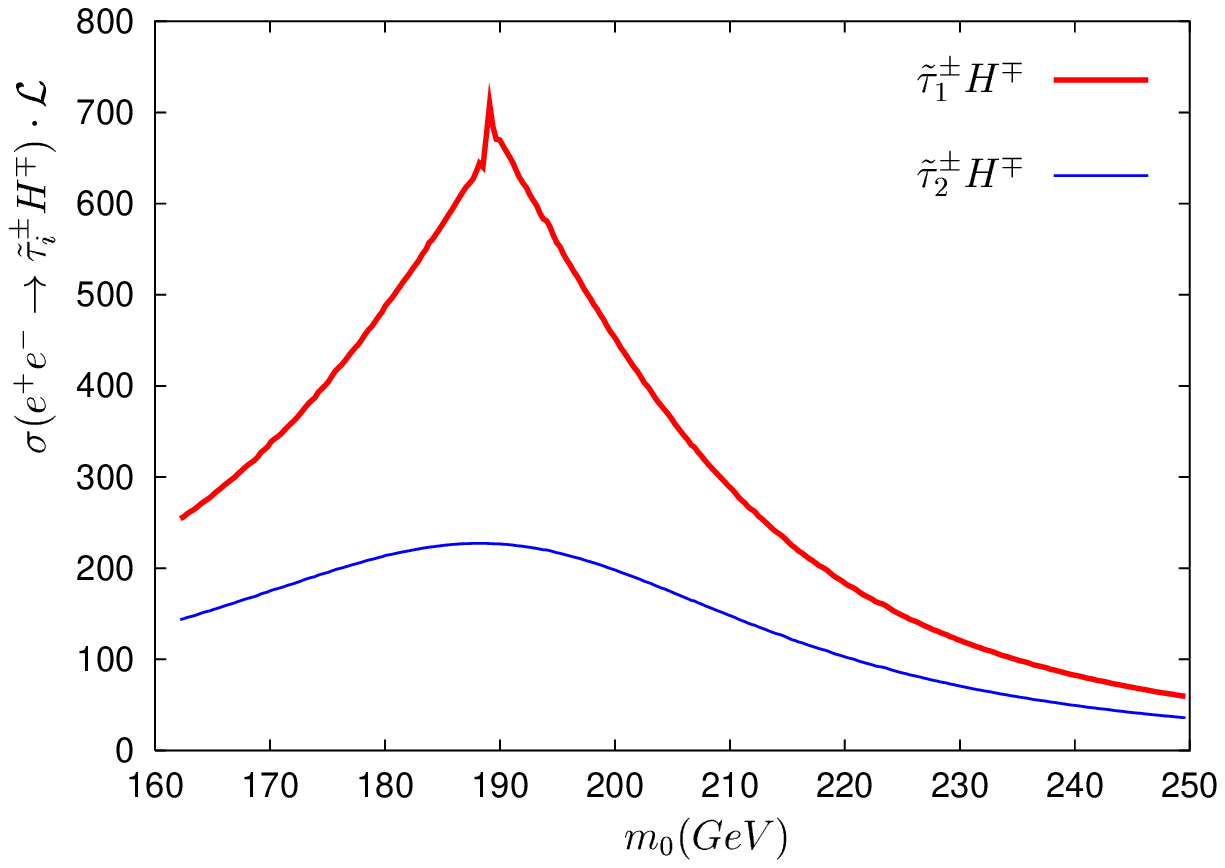}
\vskip .5cm
\caption[]{
Charged Higgs production in association with a stau multiplied by an 
integrated luminosity of $500 \,\,{\mathrm fb}^{-1}$, plotted as a function 
of the scalar mass $m_0$, for $M_{3/2}=30$ TeV, $\tan\beta=15$ GeV, 
$\epsilon_3=1$ GeV, $m_{\nu}=0.1$ eV, and $\mu<0$.
}
\label{Sm0}
\end{figure}

The dependence of these two cross sections on the scalar mass $m_0$ can 
be seen in Fig.~\ref{Sm0}, where a maximum value is achieved at $m_0=190$
GeV. Nevertheless, the maximum value for the cross section in much smaller 
than in the previous two figures, where the control parameter is 
$\tan\beta$.

\begin{figure}
\dofig{5in}{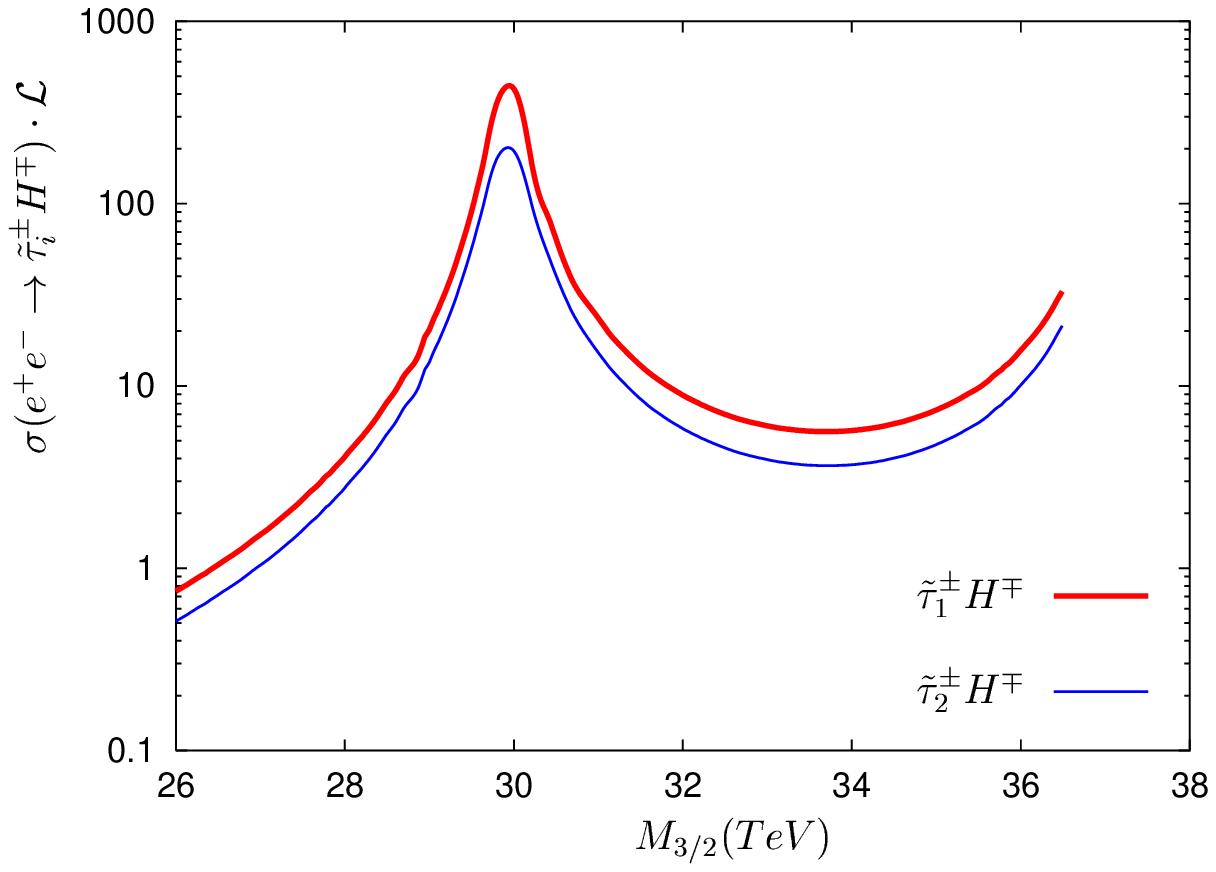}
\vskip .5cm
\caption[]{
Charged Higgs production in association with a stau multiplied by an 
integrated luminosity of $500 \,\,{\mathrm fb}^{-1}$, plotted as a function 
of $M_{3/2}$, for $m_0=200$ GeV, $\tan\beta=15$ GeV, 
$\epsilon_3=1$ GeV, $m_{\nu}=0.1$ eV, and $\mu<0$.
}
\label{Sm32}
\end{figure}

To finish this section, in Fig.~\ref{Sm32} we present the production cross
section times luminosity of a charged Higgs in association with a stau
as a function of the gravitino mass $M_{3/2}$. A sharp maximum is observed
for both cases at $M_{3/2}=30$ TeV and decreasing rapidly for smaller 
values. At higher values the cross sections start to rise again, but 
the curves end when too low values of scalar masses are reached.

\section{Extracting Supersymmetric Parameters from Observables}

As it was mentioned before, ratios of branching ratios of neutralino
decays are related to the parameters $\Lambda_i$, which are in turn
related to the atmospheric mass and angle. In this way, measurements
on neutralino decays may give information on these
parameters. Nevertheless, it is not easy to extract from neutralino
physics information on the $\epsilon_i$ parameters or on the sneutrino
vevs $v_i$. 

The situation is different in the scalar sector, and in particular, in
the charged scalar sector. As we have seen in chapter IV, mixing
angles in the Higgs stau sector depend directly upon the BRpV
parameter $\epsilon_3$ and the sneutrino vacuum expectation value
$v_3$. In addition, near degeneracy between the charged Higgs and the
stau found in AMSB enhances the associated production cross section,
making this R-Parity violating process observable.

Based on this idea, we study the possibility of extracting the
fundamental parameters of the model, and especially $\epsilon_3$ and
$v_3$, from hypothetical measurements of production cross sections and
decay rates of charged scalars. We use a $\chi^2$ method with an input
given by the AMSB model defined in eq.~(\ref{benchmark}). We randomly
generate models varying all the relevant parameters in the 
MSSM: 
\begin{itemize}
\item
Higgs sector parameters $\mu$, $\tan\beta$, $m_A$.
\item
Slepton soft mass parameters $M_L^2$, $M_R^2$.
\item
Trilinear soft mass parameters $A_{\tau}$.
\item
Gaugino masses $M_1$, $M_2$.
\item
Neutrino sector parameters $\epsilon_3$, $m_{\nu_3}$.
\end{itemize}
These parameters are free, in the sense that are not calculated with the 
AMSB boundary conditions. For the {\sl ith} model we calculate a 
$\chi^2_i$ (normalized by the number of observables) given by
\begin{equation}
\chi^2_i=\left({{\sigma_i-\sigma}\over{\delta\sigma}}\right)^2+...
\end{equation}
where $\sigma_i$ is the observable calculated with the random parameters of 
the {\sl ith} model, $\sigma$ is the observable calculated with parameters
given by the input AMSB benchmark, and $\delta\sigma$ is a projected error
in the measurement of the observable, estimated using only the statistical 
error $N\pm\sqrt{N}$. As observables we use the charged scalar masses 
in eq.~(\ref{chscalarmasses}) (with a projected error of $1\%$), the 
production cross sections in eq.~(\ref{CrossSections}), and branching ratios
of charged scalars decaying into charginos, neutralinos, and leptons.

\begin{figure}
\centerline{\protect\hbox{\epsfig{file=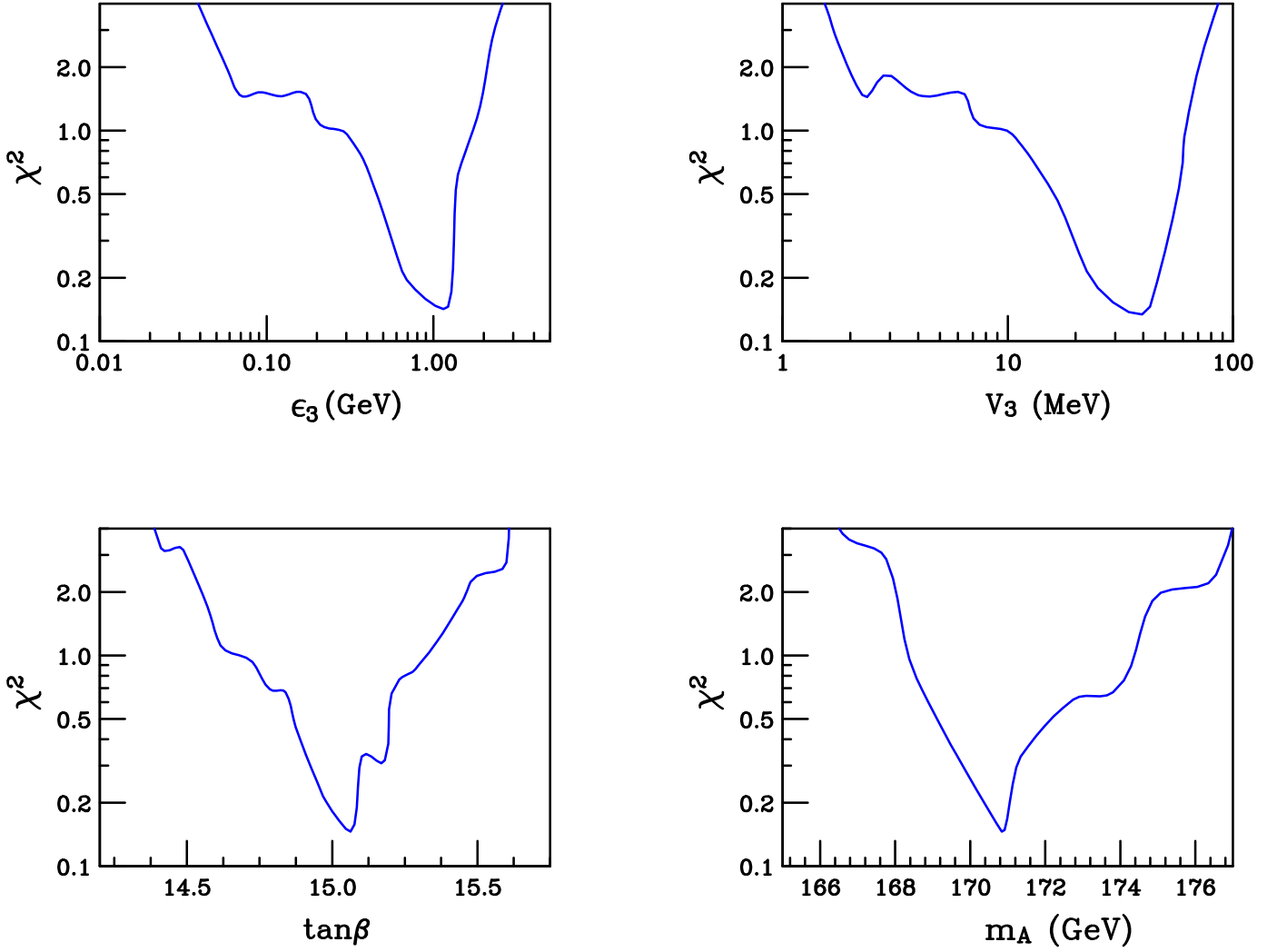,width=0.85\textwidth,angle=90}}}
\vskip .5cm
\caption[]{
Normalized $\chi^2$ lower bound as a function of the parameters 
(a) $\epsilon_3$, (b) $v_3$, (c) $\tan\beta$, and (d) $m_A$. The boundaries 
indicated by $\chi^2=1$ defines the expected error in the determination 
of these parameters.
}
\label{1_4_chi2}
\end{figure}

\begin{figure}
\centerline{\protect\hbox{\epsfig{file=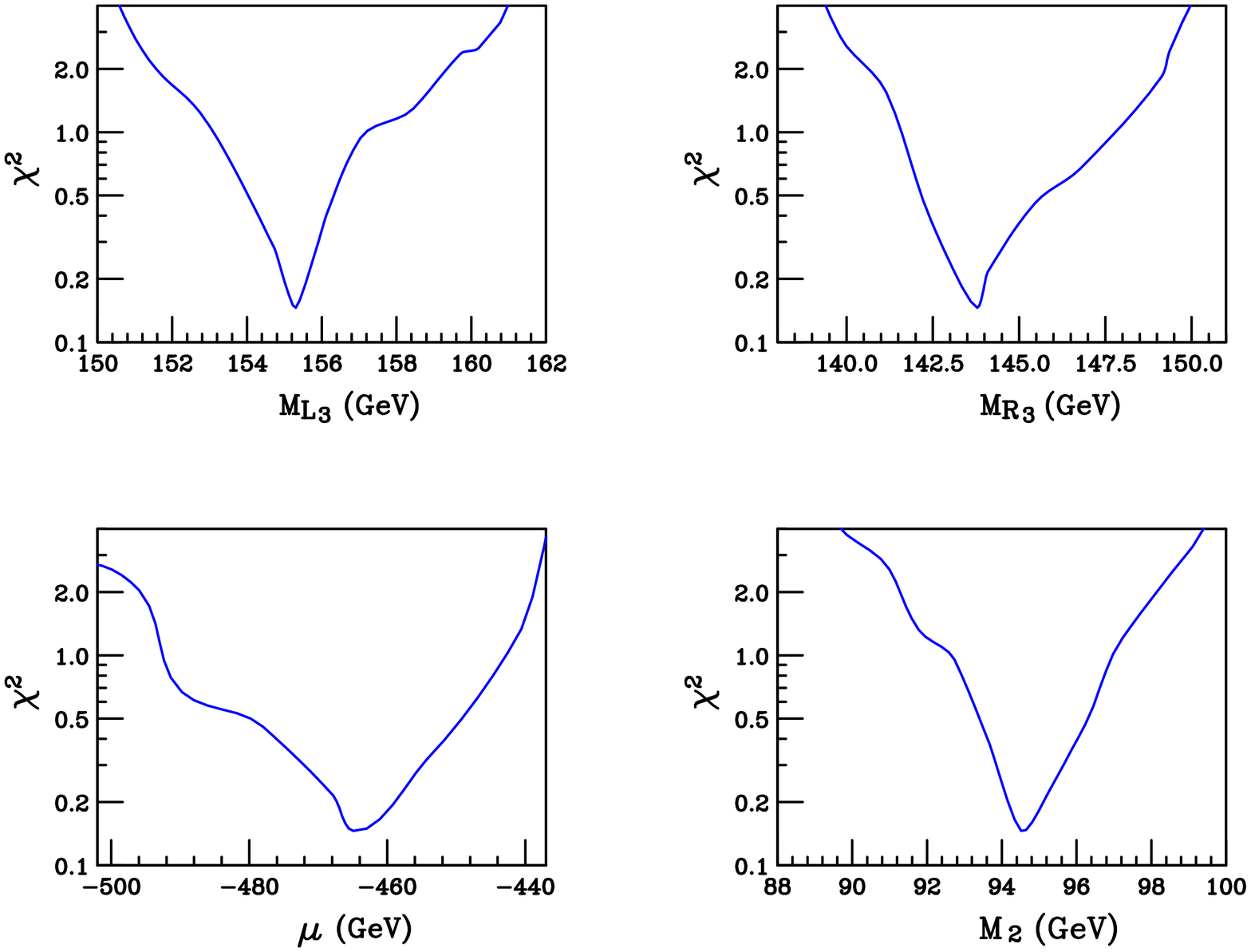,width=0.85\textwidth,angle=90}}}
\vskip .5cm
\caption[]{
Normalized $\chi^2$ lower bound as a function of the parameters 
(a) $M_{L_3}$, (b) $M_{R_3}$, (c) $\mu$, and (d) $M_2$. The boundaries 
indicated by $\chi^2=1$ defines the expected error in the determination 
of these parameters.
}
\label{2_4_chi2}
\end{figure}

{}From a large sample of models we select 2000 of them with normalized 
$\chi^2<2$ and plot the $\chi^2$-distributions in Figs.~\ref{1_4_chi2}
and \ref{2_4_chi2}. The most interesting distributions for us are the 
corresponding to $\epsilon_3$ and $v_3$ in Fig.~\ref{1_4_chi2}. In this 
figure we see that a clear upper bound on these parameters can be set, 
in addition to a less clear lower bound. This is an important achievement
considering the difficulties in extracting values for these parameters 
from neutralino decay measurements. From the rest of the 
$\chi^2$-distributions we learn that a reasonable determination of
$\tan\beta$, $m_A$, $M_{L_3}^2$, $M_{R_3}^2$, $\mu$, and $M_2$ can be made.
No useful information can be obtained for the parameter $A_{\tau}$, and
only limited information for $M_1$ (which we do not show).

\begin{figure}
\dofig{5in}{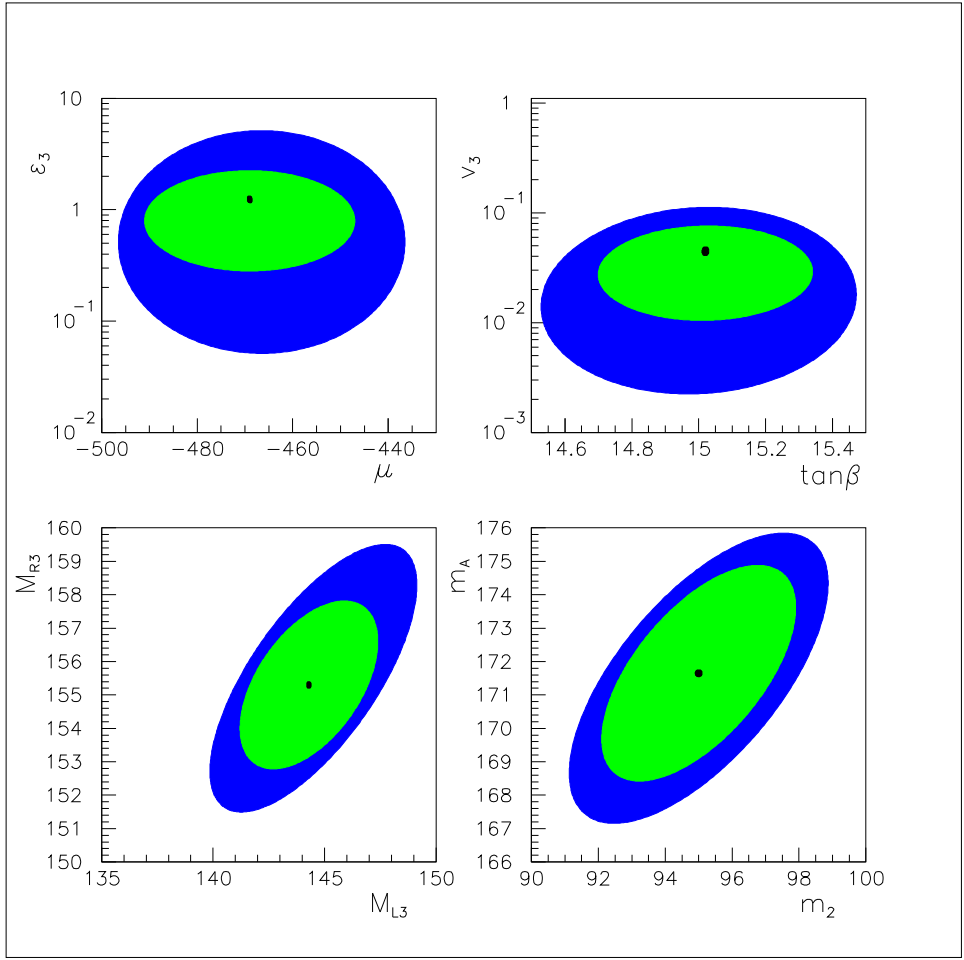}
\vskip .5cm
\caption[]{
Regions of parameter space where normalized $\chi^2\le 1$ are shown in 
green (light gray). For comparison, also shown are the regions where 
$\chi^2\le 2$ in
blue (dark gray).
}
\label{chi2_sca}
\end{figure}

In Fig.~\ref{chi2_sca} we have regions in different planes of parameter
space where normalized $\chi^2<1$ (light grey, or green), which tells us 
about the error in the determination of the corresponding parameter. 
For comparison we also show the regions with $\chi^2<2$ (dark grey, or 
blue). From this figure we extract the output for each parameter from 
the $\chi^2$ analysis, the error in its determination, and compare them
with the input values from our benchmark. This comparison is shown in
Table \ref{tab:cases}. The determination of all the parameters of the 
R-Parity conserving MSSM directly involved in the Higgs/slepton sector
is very good, with a few percent of error. The fact that we can 
set an upper and lower bound on the R-Parity violating parameters 
$\epsilon_3$ and $v_3$ is good in itself, although with large errors.

The enhancement of the charged scalar mixing angles due to near degeneracy
will modify the prediction of the solar mixing angle and mass scale. This 
possibility will be studied in a further work where BRpV in three 
generations is included.

\newpage
\section{Conclusions}

Supersymmetric models with bilinear R-Parity violation predict neutrino 
masses and mixing angles which agrees with experimental data on solar
and atmospheric neutrinos. It has been shown that the supergravity version
of this model is testable at colliders via neutralino decays, when this 
particle is the LSP, and that information on the parameters $\Lambda$ can 
be obtained. In this article we show that BRpV embedded into an
Anomaly Mediated Supersymmetry Breaking model can also be tested at 
colliders with processes not necessarily related to LSP decay. In fact,
we show that in cases of near degeneracy between charged Higgs boson and 
staus, the charged scalar mixing is enhanced, as well as the charged Higgs 
production in association with staus. The end result is that it is possible 
to determine the values of the parameters $\epsilon_3$ and $v_3$ from
measurements of charged scalar masses, production cross sections, and decay
rates. 

\vskip 0.5cm
\noindent
{\bf Note Added:}
When this article was being written we received a related work where
neutrino physics is probed at colliders with charged slepton decays in
Supergravity when the slepton is the LSP\cite{HPRV}. The ideas presented
in their article are in agreement with ours.

%
\acknowledgments
This research was supported in part by Conicyt grant No.~1010974, and 
DIPUC grant No.~2000/08E.

%

%

\newpage
%
%
\begin{table}
\begin{center}
\caption{Parameter determination from charged Higgs and stau associated 
production and decay for a given case study (benchmark). All parameters 
are expressed in GeV, except for $\tan\beta$.
}
\bigskip
\begin{tabular}{lccccc}
\hline
parameter & input & output  & error & percent  \\
\hline

$\epsilon_3 $ & 1.0   & 1.24   & 0.96    & 77     \\
$v_3        $ & 0.035 & 0.045  & 0.035   & 78     \\
$\tan\beta  $ & 15.0  & 15.0   & 0.3     &  2     \\
$\mu        $ & -466  & -469   & 22      &  5     \\
$M_{L_3}    $ & 155.8 & 155    & 3       &  2     \\
$M_{R_3}    $ & 144.5 & 144    & 3       &  2     \\
$m_A        $ & 171.5 & 172    & 3       &  2     \\
$M_2        $ & 95.4  & 95     & 3       &  3     \\
\hline
\label{tab:cases}
\end{tabular}
\end{center}
\end{table}

\newpage
%

%
\newpage

\end{document}